\documentclass[preprint]{elsarticle}
\usepackage[utf8x]{inputenc}
\usepackage{multirow,graphicx,epsfig,color}

\begin{document}

\begin{frontmatter}
\title{DFT study of noble metal impurities on TiO$_{\mathbf{2}}$(110)}

\author[mete]{Ersen Mete\corref{cor1}}
\ead{emete@balikesir.edu.tr}
\author[gulseren]{O\u{g}uz G\"{u}lseren}
\author[ellialtioglu]{\c{S}inasi Ellialt{\i}o\u{g}lu}

\address[mete]{Department of Physics, Bal{\i}kesir University, Bal{\i}kesir
10145,Turkey}
\address[gulseren]{Department of Physics, Bilkent University, Ankara 06800, Turkey}
\address[ellialtioglu]{Department of Physics, Middle East Technical University, Ankara
06800, Turkey}

\cortext[cor1]{Corresponding author.}

\begin{abstract}
Atomic and electronic structures of TiO$_2$(110) surface with possible
adsorptional, substitutional and interstitial Au or Pt elemental
impurities at full and one-sixth monolayer concentrations were investigated
by density functional theory calculations using the projector augmented wave
approach within the plane wave method. Relative thermodynamic stabilities of
such phases have been discussed by means of their surface free energies. Our
results suggest that tunable photocatalytic activity can be achieved on Pt atom
admixed rutile (110) surface at low coverages. 
\end{abstract}


\end{frontmatter}

\section{Introduction}

Titanium dioxide (TiO$_2$) attracts an increased interest as a catalyst
support. Gold nanoparticles on titania were shown to exhibit
remarkable catalytic activity~\cite{haruta}. Another important metal-on-oxide
system is platinum incorporated TiO$_2$ which finds applications as catalysts
and gas sensors~\cite{henrich}. Pt/TiO$_2$ system has drawn further attention
because of its photocatalytic activity towards water
decomposition~\cite{linsebigler}. In recent studies, these noble metals
supported on TiO$_2$ has shown to perform highly efficient catalysis under
solar light irradiation~\cite{varghese,rosseler,du,choi}.

As a wide-gap semiconductor, the rutile TiO$_2$(110) surface is considered
to be the generic model system for oxide surfaces. The (110) surface of
rutile structure has the lowest surface energy among the other
facets~\cite{oviedo1,oviedo2,slater,sensato,beltran}. It is reducible by
surface oxygen vacancy creation or metal incorporation which attracts a great
deal of interest for fundamental study of photo- and heterogeneous
catalysis~\cite{litter,hangfeldt,gratzel01,khan}, functional ultrathin
films~\cite{chen06,finetti} and dielectrics~\cite{wu,griffin}. Understanding
of the properties of metal--metal oxide interface can provide important
insights into the applications of real catalysts.

Bonding of gold on titania has been studied to shed light on the effect of
Au--support adhesion on its catalytic behavior both experimentally\cite{l_zhang,
chen04,maeda, benz,locatelli,matthey,tong} and theoretically~\cite{z_yang,
vijay,wang,pillay1,iddir05,pillay2,o_maeda1,o_maeda2, okasawa,chretian,marri,
pabisiak,f_yang}. Similarly, the adsorption properties of Pt have also been
investigated by many experiments~\cite{linsebigler,steinruck,sasahara1,
sasahara2,iliev,park,isomura} and by a few theoretical studies~\cite{thien_nga,
iddir05,veysel}. Although the nobel metal-enhanced catalytic activity of rutile
TiO$_2$ has been reported by experiments, the electronic properties of such
systems as the originating factor are still not well known.

Moreover, the composition of nobel metals like Au and Pt to TiO$_2$ support can
be interstitial or substitutional as well as being adsorptional. Indeed,
experiments showed that Pt atoms can thermally diffuse into TiO$_2$ lattice
under oxidizing atmosphere~\cite{m_zhang}. Futhermore, these diffused Pt atoms
can substitute Ti$^{4+}$ when oxidized to Pt$^{2+}$ or they form interstitials
inside. Such metal impurities are known to greatly influence the electronic and
catalytic properties of the combined system.

In this paper, we studied the structural and electronic properties of
adsorptional, substitutional, and interstitial Au or Pt impurities on rutile
TiO$_2$(110) surface at high (1 ML) and at low (1/6 ML) concentrations.
Relative thermodynamic stabilities of such impurity phases have also been
discussed. Our primary aim is to elucidate the effect of Au(Pt)
incorporation on the electronic structure of the titania support at the
fundamental level. In this sense, we give emphasis on the electronic behavior
-- especially on the photocatalytic activity -- as a result of the bonding
characteristics of such an incorporation rather than investigating the
formation of such impurities.

\section{Method}

The calculations have been performed by the density functional theory (DFT)
implementation of the VASP~\cite{vasp} code. Exchange--correlation energy
has been approximated by the gradient corrected Perdew--Burke--Ernzerhof
(PBE96) functional~\cite{pbe}. We used projector augmented waves (PAW)
approach~\cite{paw1,paw2} with a plane-wave basis up to a cutoff of 400 eV.

We considered the bulk terminated (1$\times$1) and (3$\times$2) supercells
as the slab models for high (1 ML) and low (1/6 ML) metal coverages,
respectively. Bulk terminated rutile TiO$_2$(110)-(1$\times$1) and
TiO$_2$(110)-(3$\times$2) surfaces have been modeled by a symmetric slab of
7 TiO$_2$ trilayers separated by $\sim$15 {\AA} of vacuum region. Each trilayer
consists of a central O--Ti--O plane and 2 oxygen atoms placed symmetrically
above and below this plane.

For the geometry optimization calculations, the Brillouin Zones of (1$\times$1)
and (3$\times$2) supercells were sampled with a 8$\times$5$\times$1 and
2$\times$2$\times$1 $k$-point meshes, respectively. In all calculations, the
full relaxation has been performed using conjugate-gradient algorithm based on
the reduction of the Hellmann--Feynman forces on each atom to less than 0.01
eV{\AA}$^{{\rm -}1}$. We used much denser grids for the computations of band
structures and densities of states (DOS).

We calculated the binding energy of adsorbate, M, by
$$
E_b=(E_{{\rm M/TiO}_2}-E_{{\rm TiO}_2}-2E_{\rm M})/2
$$
where $E_{{\rm M/TiO}_2}$ is the total energy of the computation cell
involving the slab and the atomic impurities, M, $E_{{\rm TiO}_2}$ is
that of the defect-free stoichiometric slab, and $E_{\rm M}$ is the
energy of an isolated M (Au, Pt) atom calculated in its electronic ground
state.

The formation energies of defects in the form of metal atom impurities
on (1$\times$1) and (3$\times$2) surfaces has been calculated (as previously
described in detail\cite{mete1,mete2}) by
$$
E_f=(E_{{\rm M/TiO}_2}-E_{{\rm TiO}_2}-\Delta m\,\mu_{\rm M}+\Delta n\,\mu_{\rm Ti})/2A
$$
where $\Delta m$ and  $\Delta n$ are the differences in the number of adsorbate
M atoms and surface Ti atoms from the reference stoichiometric TiO$_2$ slab,
respectively. The chemical potentials, $\mu_{\rm M}$, were taken from their
reference bulk values of $-3.270$ eV for Au and of $-6.017$ eV for Pt, representing
their most stable solid phases accessible. In other words, we assume that the
surface is in thermodynamic equilibrium with ccp bulk Au or Pt. By the same
token, the surface layer must be in equilibrium with the rutile TiO$_2$ bulk
which comes into contact with. This physical requirement implies
$\mu_{\rm Ti}+2\mu_{\rm O} = \mu_{{\rm TiO}_2}$ restricting, $\mu_{\rm Ti}$
within an interval of allowed values. Chemical potential of Ti can be as high
as that of its bulk which defines an upper boundary referring to Ti-rich
conditions. On the other hand, molecular oxygen defines the most stable phase
for $\mu_{\rm O}$ so that $\mu_{\rm O} = \frac{1}{2} E_{\rm{O}_2}$ referring to
O-rich conditions. This choice, therefore, defines the minimum value of
$\mu_{\rm Ti}$ through the thermodynamic equilibrium condition,
$\mu_{\rm Ti}+ 2\mu_{\rm O} = \mu_{{\rm TiO}_2}$.

\section{Results and Discussion}

Rutile TiO$_2$ has a direct band gap of 3.03 eV~\cite{pascaul} at $\Gamma$
corresponding to UV optical response. Our calculated value is 1.85 eV because
of the well known underestimation of GGA functionals due to insufficient
cancellation of the self-interaction energy. The band gap can greatly be
corrected by employing many-body perturbative corrections up to first order in
the screened Coulomb potential, W, called as the GW approximation. A better
description of both electronic and optical spectra can be obtained by applying
Bethe--Salpeter equation (BSE) including excitonic effects.
Such a quasiparticle treatment for the correlation energy starting from the
DFT spectrum (Kohn--Sham eigenvalues and wavefunctions) of TiO$_2$ bulk
material has been reported to give scissors-like correction to unoccupied
states without noticeable change in the band
dispersions~\cite{chiodo,kang}. Therefore, descriptions of electronic
properties based on pure DFT results can be made as far as the band structures
are concerned. After all, many-body perturbative approach for a supercell
consisting of TiO$_2$(110)-(3$\times$2) slab with 7 trilayers and a vacuum
region of $\sim$15 {\AA} is computationally very expensive, if not impossible
even at the GW level.

Another important factor is that the surface energy tend to converge with
increasing number of trilayers with an odd--even oscillation similar to those
reported previously~\cite{ramamoorthy,x_wu,bredow,thompson,hameeuw,labat,kowalski}.
After testing the convergence of surface energetics as a function of the slab
thickness we have chosen the seven trilayer model as a compromise between the
accuracy and the computational cost. The present work not only uses a larger cell
(e.g. 252 atoms for (3$\times$2) cell) compared to the previous studies but also
puts forward the electronic band structures of TiO$_2$(110) surface with
Au(Pt) atomic impurities at 1 and 1/6 ML coverages. This allows us to discuss
the effects of such a metal incorporation on the electronic structures
specifically in the gap region to get a better understanding.

\subsection{Au(Pt)/TiO$_2$(110)-(1$\times$1)}

\begin{figure}
\begin{center}
\epsfig{file=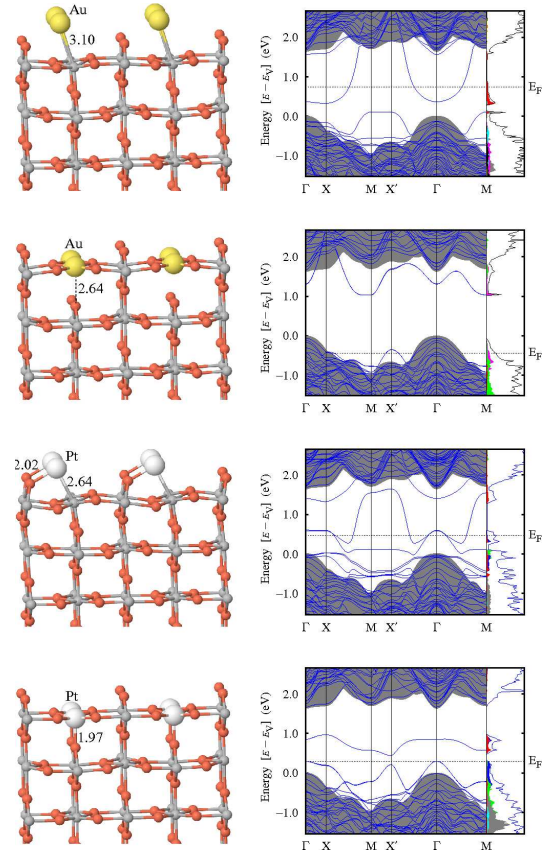,width=7cm,clip=}
\caption{Relaxed geometries (top three trilayers) of the
Au(Pt)/TiO$_2$(110)-(1$\times$1) systems are presented on the left. Their
band-gap states along with the corresponding bulk projected band structures
(shaded areas) and DOS are shown on the right, next to them. Gray and dark gray
(red in color) balls denote Ti and O atoms, respectively. Some key distances
between Au(Pt) and (sub)surface atoms are depicted (in {\AA}).}
\label{fig1}
\end{center}
\end{figure}

At 1 ML coverage, Au binds to fivefold coordinated surface Ti (Ti5c) with a
bond length of 3.10 {\AA} [$d_{\rm Au-Ti}$ in Table~\ref{table1}] tilted by
14$^\circ$ toward the nearest-neighbor (nn) bridging oxygen (O2c) as shown
in Fig~\ref{fig1}. Well-ordered Au monolayers where gold being atop Ti5c on
(1$\times$1) surface has also been verified by several experiments~\cite{wang,
chen04,benz}. Single Au adsorbates are separated from each other by 2.96 {\AA}
along [001] direction. Our calculated gas phase Au dimer length is 2.53 {\AA}
in agreement with Hakkinen \textit{et al.}'s theoretical result of 2.54
{\AA}~\cite{hakkinen}. Hence, Au adatoms on (1$\times$1) surface form linear
metallic chains parallel to the oxygen rows where the interatomic distances are
slightly longer than the gold dimer length. This indicates a strong Au--Au
coordination. Projected DOS analysis shows that the peak corresponding to the
half filled gap state is mainly due to Au 5$d$ electrons. It exhibits high
dispersion along X--M and X$'$--$\Gamma$--M reflecting the strength of the
bonding along 1D gold chain. It couples to the conduction band (CB) of TiO$_2$
and is almost flat between $\Gamma$ and X (reciprocal to [1\={1}0] direction)
indicating a weak adsorption. Yang \textit{et al.} found the binding energy of
1 ML Au on the relaxed TiO$_2$(110) to be 1.49 eV at Ti5c site with a Au--Ti
bond length of 2.66 {\AA} through full potential linearized augmented
plane-wave (FLAPW) calculations~\cite{z_yang}. We calculated the binding energy
(BE) of an Au atom on the stoichiometric (1$\times$1) rutile surface to be 1.38
eV. In fact, this BE does not reflect bare metal--substrate interaction, it
substantially involves interaction of gold with its periodic images on
(1$\times$1) cell causing a significant increase in adsorption energy. Indeed,
when we calculated the BE on (3$\times$2) surface it reduces to 0.40 eV  where
Au adatoms are separated from each other by 8.9 {\AA} along [001] and by 13.1
{\AA} along [1\={1}0]. Therefore, Au weakly binds to the surface. Moreover,
calculating the BE's at all possible adsorption sites we obtained an almost
flat potential energy surface similar to that of Iddir \textit{et al.}~\cite{iddir05}. 
This suggests that gold can diffuse in all directions over the
surface in agreement with experimental findings~\cite{maeda,locatelli,matthey}.

\begin{table}[bt]
\begin{center}
\caption{Calculated values for the M/TiO$_2$(110) systems
(M=Au, Pt): work function and Fermi energy relative to bulk valence band top (in eV),
as well as M--O and M--Ti distances (in \AA) for each model.}
\label{table1}
\begin{tabular}{lcc|cl|cc}\hline\hline
\raisebox{4.5mm}{\mbox{}} \raisebox{-3mm}{\mbox{}}Slab & \multicolumn{2}{c|}{System} &
~~$\Phi$~~ & ~~~$E_{\rm F}$~ & $d_{\rm M-O}$ & $d_{\rm M-Ti}$ \\ \hline
(1$\times$1) & \multicolumn{2}{c|}{clean} & 7.22 & ~~0.00 & --- & --- \\ \hline
& \multirow{2}{*}{Au} & on & 6.16 & ~~0.76 & 2.74 & 3.10 \\
& & subst & 7.39 & $-0.40$ & 1.97 & --- \\[-4mm] \\ \hline
& \multirow{2}{*}{Pt} & on & 5.00 & ~~0.49 & 2.02 & 2.64 \\
& & subst & 6.78 & ~~0.33 & 1.96 & --- \\[-4mm] \\ \hline\hline
(3$\times$2) & \multicolumn{2}{c|}{clean} & 7.26 & ~~0.00 & --- & --- \\ \hline
& & on & 4.97 & ~~1.75 & 2.05 & 3.59 \\
& Au & subst & 7.56 & $-0.05$ & 2.03 & --- \\
& & in & 5.69 & ~~1.80 & 2.18 & 2.75 \\ \hline
& & on & 6.23 & ~~0.63 & 1.96 & 2.41 \\
& Pt & subst & 7.39 & ~~0.03 & 2.01 & --- \\
& & in & 6.97 & ~~0.46 & 1.99 & 2.70 \\ \hline\hline
\end{tabular}
\end{center}
\end{table}

Gold substitution for Ti5c at 1 ML causes considerable distortion on the
surface morphology and on the electronic structure as shown in the second row
of Fig.~\ref{fig1}. The distance between Ti5c and the oxygen beneath it
extends from 1.83 {\AA} to 2.64 {\AA} by Au substitution. Metal dopant
interacts with four nn threefold coordinated basal oxygens (O3c) much
weaker than Ti does. An explanation might be that the valence of Au,
$6s^{1}$, compared to the valence of Ti, $3d^34s^1$, imposes an electron
deficiency. As a result, a half filled Au--O3c driven state appear within
the VB setting the Fermi energy below the bulk projected VBM of TiO$_2$.
Additionally, an unoccupied impurity state falls in the gap with a coupling
to the CB at about X.

Low energy Pt adsorption site is above Ti5c tilted by 24.7$^\circ$ toward the
nn O2c similar to single Au adsorption case but closer to the surface. This
site is also referred as the hollow site since it is atop the middle point
between two basal oxygens~\cite{iddir05}. This prediction slightly disagrees
with experimental adsorption site atop Ti5c, probably due to a difference in
the theoretically predicted and experimentally measured amounts of charge
transfer from Pt to TiO$_2$~\cite{sasahara1}. Pt adatom causes noticeable
lattice distortions up to the second trilayer (Fig.~\ref{fig1}). Unlike the
gold case, Pt adatom interacts strongly with rutile (110) surface giving a BE
of 2.79 eV on (1$\times$1) cell. It decreases to 2.17 eV on (3$\times$2) cell.
Clearly, this lowering is much smaller than that of the Au adatom case, since
the contribution from the interaction of Pt with its periodic images is smaller.
This is because of the fact that Pt--Pt separation along [001] on (1$\times$1)
surface is considerably larger than the Pt dimer length of 2.33 {\AA} due to
the lattice parameter ($c$=2.96 {\AA}) of the bulk TiO$_2$. Moreover, Pt ML
reduces the work function of the clean surface by 2.22 eV (Table~\ref{table1}),
more than Au does due to larger amount of charge transfer from metal to
substrate, indicating a stronger binding. Hence, a uniform Pt deposition is
relatively more probable than an Au adlayer formation. In fact, Steinr\"{u}ck
\textit{et al.} reported that Pt deposition happens to cause a uniform coverage
on the surface at low temperatures ($<$\,160\,K). Higher substrate temperatures
lead to Pt islands by increasing Pt diffusion probability~\cite{steinruck}.
Pt--O2c and Pt--Ti5c bonds bring two impurity states that disperse strongly
across the band gap. Their flat-like dispersion along $\Gamma$--X reveals the
weakness of Pt--Pt interaction along [1\={1}0]. Fermi energy crosses these
states leading to metallization. Pt induced distortions, localized to surface
layer, bring states at and around the VBM. An upper lying empty impurity state
couples to the CB near and between M and X$'$ modifying its edge.

\begin{figure}[tbh]
\begin{center}
\epsfig{file=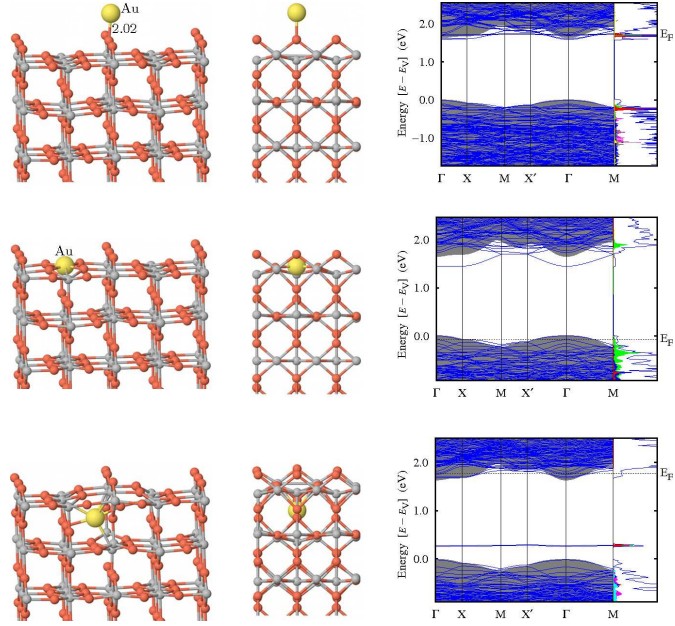,width=8.8cm,clip=}
\caption{Top three trilayers of fully relaxed
Au/TiO$_2$(110)-(3$\times$2) systems along [1$\bar{1}$0] and [001] directions
for adsorptional (on), substitutional (subst) and interstitial (in) cases
are shown. Relevant energy band diagrams are presented with DOS (in the
rightmost panel) for each of these atomic structures.}
\label{fig2}
\end{center}
\end{figure}

Substitutional Pt on (1$\times$1) surface does not distort the surface layer as
much as an Au dopant does (at the bottom row of Fig.~\ref{fig1}). The distance
between Pt and subsurface O is 1.97 {\AA} that is slightly extended with respect
to Ti--O bond length of 1.83 {\AA}. Having fourfold coordination with basal
oxygens, Pt dopant modifies the VBM edge by inducing a gap state that elevates
the Fermi energy by 0.33 eV at $\Gamma$ relative to the VBM of the clean surface.
Another unoccupied state appears a $\sim$0.1 eV above the Fermi energy
leading to a narrow-gap (indirect between $\Gamma$ and X$'$) semiconducting
system.

\subsection{Au(Pt)/TiO$_2$(110)-(3$\times$2)}

Minimum energy adsorption site for single Au on the (3$\times$2) cell with
21 layers has been found to be at above bridging oxygen, O2c as shown in
Fig.~\ref{fig2} in agreement with previous theoretical
results~\cite{o_maeda2,vijay,matthey}. Another preferential site is the the
hollow site~\cite{pillay1,iddir05,chretian}. These two configurations differ
insignificantly in their total energies and are both experimentally
verified~\cite{tong}. Campbell \textit{et al.} reported Au monomer adsorption
energy to be 0.43 eV by calorimetric measurements~\cite{campbell}. Vijay
\textit{et al.} found that Au binds to an O2c or to a Ti5c atom (with a tilting
toward an O2c) weakly by 0.6 eV. We calculated the binding energy of Au on the
(3$\times$2) cell to be 0.40 eV which indicates a flat-like potential energy
surface. Indeed, the diffusion barriers were found to be so low that Au atoms
already diffuse at room temperature (RT) both in the [001] and in the [1\={1}0]
directions~\cite{matthey,worz,del_vitto,iddir05}.

\begin{figure}[htb]
\begin{center}
\epsfig{file=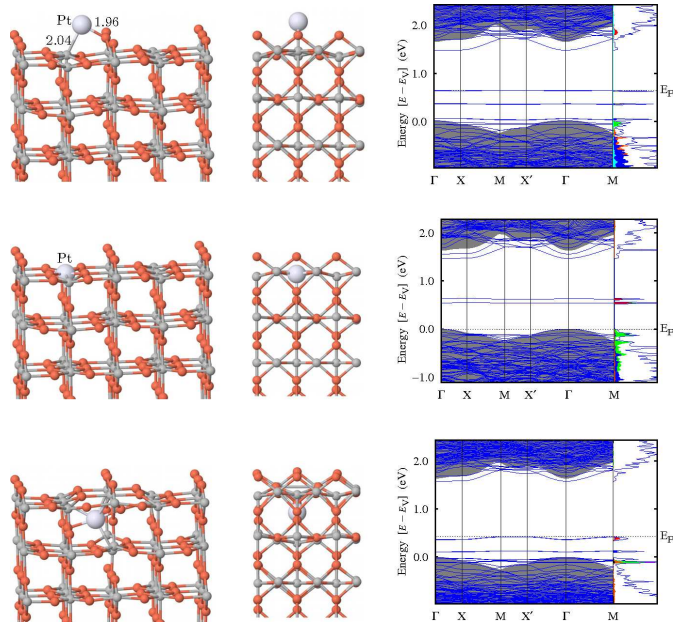,width=8.8cm,clip=}
\caption{Atomic structures of Pt/TiO$_2$(110)-(3$\times$2) systems along
[1$\bar{1}$0] and [001] directions for adsorptional (on), substitutional
(subst) and interstitial (in) cases. Relevant energy band diagrams and
DOS are presented on the right for each of them.}
\label{fig3}
\end{center}
\end{figure}

On stoichiometric TiO$_2$(110) surface Au is reported to exhibit a
quasi-two-dimensional growth at low concentrations at low temperatures.
\cite{l_zhang,maeda} As the deposition rate or temperature increases Au
starts to form three dimensional islands. Formation of Au clusters on
the surface indicates a very weak metal--substrate interaction. On the
other hand, strong binding can be achieved at O2c vacancies. For example,
Benz \textit{et al.} deposited Au atoms on titania surface forming monatomic
Au centers~\cite{benz}. Following experiments by Tong \textit{et al.} revealed
that Au binding can be broken by a hydroxyl group forcing Au out of the
adsorption site in the presence of water.

Electronically, single Au adsorbate causes metallization due to an unpaired
$5d$ electron. Moreover, it gives a state just around the CBM and modifies the
CB edge. Okazawa \textit{et al.} determined the work function of TiO$_2$(110)
surface with low gold coverage to be $\sim$5.3 eV where our value is 4.97 eV.
On the other hand, their clean surface value of 5.4 eV is nearly 2.2 eV
smaller than ours. The discrepancy can be addressed to possible existence of
oxygen vacancies on the sample because such a reduction was shown to cause
2 eV drop in the work function by Vogtenhuber \textit{et al.}~\cite{vogtenhuber1}.
Our clean surface value rather agrees with theirs of 7.16 eV and also with an
experimental result of 6.83 eV as reported in Ref.~\cite{vogtenhuber2}.

\begin{figure}[htb]
\begin{center}
\epsfig{file=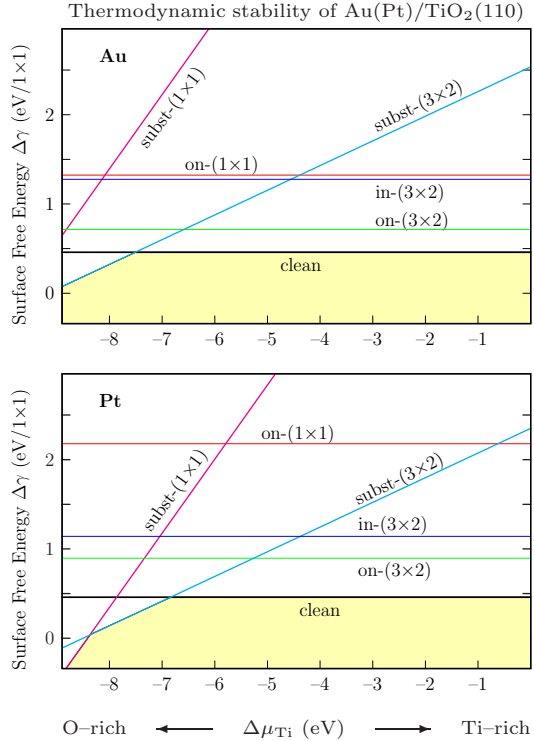,width=7cm}
\caption{Calculated formation energies of Au(Pt)/TiO$_2$(110)
systems as a function of the chemical potential of Ti. \label{fig4}}
\end{center}
\end{figure}

When Au substituted for an in-plane Ti5c, it distorts nn oxygens similar
to (1$\times$1) case, only locally. Furthermore, the overall band structure
is very similar too. Due to charge deficiency, Fermi energy falls in the
VB of TiO$_2$ giving rise to metallization. Clearly, in substitutional cases
gold also raises the work function by 0.17 on (1$\times$1) and by 0.30 eV
on (3$\times$2) cells due to valence electron deficit.

We found that precious metal atom implantation into TiO$_2$(110) subsurface
interstitial sites at 1 ML concentration is unstable. However, our calculations
suggest that metal atoms are possible as interstitials in (3$\times$2) cell,
probably at elevated temperatures and/or under oxidizing conditions in agreement
with the experimental findings~\cite{m_zhang,wendt}. Relevant surface energy
diagrams are presented in Fig.~\ref{fig4} where the formation of Au in
(3$\times$2) phase is about 0.75 eV higher than that of the clean surface.
Single Au implanted cell appears to be metallic just like the previous cases. A
flat going impurity state lies $\sim$0.3 eV above the VBM of TiO$_2$.

Adsorbate-induced modification of the substrate morphology is more pronounced
for Pt on rutile (110) relative to the Au case. For instance, Pt pulls nn
Ti5c and O2c toward itself at the expense of extending their bonds with the
lattice. This indicates relatively stronger binding with 2.17 eV adsorption
energy at the hollow site (see Fig.~\ref{fig3}) in agreement with the BE of 2.14
eV reported by Iddir \textit{et al.} for Pt on (3$\times$2) with 4 trilayers. This
site differs with a tilting angle of $\sim$27$^\circ$ from experimentally
assigned adsorption atop Ti5c atom~\cite{sasahara1}. The disagreement has been
addressed to the difference in theoretically predicted and experimentally
estimated amounts of charge on Pt. Experiments confirm the enhancement of
photocatalytic activity of titania surfaces by platinization~\cite{park,iliev}.
This can be partly explained by the availability of defect driven gap states
that increase transition probabilities. For example, the excess charge localized
on the Pt adsorbate causes three flat going states in the band gap, which can
also be seen from their strong DOS peaks. These states lie at 0.04, 0.35, and
0.63 eV above the VBM at $\Gamma$. Moreover, surface states appear near band
edges due to local distortions. A significant band gap narrowing of 0.84 eV is
predicted resulting from upper lying flat-like occupied state (0.63 eV above
VBM) and an unoccupied surface state falling into the gap from the CB (0.21 eV
below CBM at $\Gamma$).

Relaxation of the clean surface cell causes Ti5c atoms to sink. But, substituted
Pt stays levelled with in-plane oxygens (Fig.~\ref{fig3}). It slightly pulls
the subsurface O3c up. Pt mediates less distortion to the lattice in comparison
to the gold case. Moreover, Pt substitution on (3$\times$2) cell is semiconducting
whereas a single gold dopant leads to metallization. In addition, we found that
this phase is thermodynamically more stable than the clean surface under
oxidizing conditions (Fig.~\ref{fig4}b). Choi \textit{et al.} has recently shown
that Pt at low doping level significantly enhance photocatalytic activity of
rutile TiO$_2$~\cite{choi}. This must be related to the impurity states. Pt
substitute increases the work function by $\sim$0.13 eV indicating a charge
transfer to the lattice that leaves behind empty valence levels on Pt atom.
Therefore, Pt substitution on (3$\times$2) cell mainly brings unoccupied gap
states. Two of which are flat and lie just 0.54 and 0.62 eV above the VBM.
A number of impurity induced states fall into gap from the CB that show
bulk-like dispersion. Together with these states significant band gap
narrowing mimics the observed increase in photoreactivity.

Zhang \textit{et al.} reported that Pt can substitute Ti$^{4+}$ under oxidizing
conditions and can also thermally diffuse into TiO$_2$ substrate~\cite{m_zhang}.
Therefore, we considered Pt at the interstitial site as shown in the last row
of Fig.~\ref{fig3}. We calculated the formation of such an interstitial phase is
$\sim$0.65 eV relative to that of the bulk terminated surface
(Fig.~\ref{fig4}b) in line with the experiment. Although it distorts the
lattice more than the previous cases,  its effect on the energy band
structure appears to be comparable to the adsorption phase due to similar
bonding characteristics. Pt interstitial brings three occupied states
with sharp DOS peaks implying charge localization around Pt. One of these
couples to the VB of the clean surface. Relative to the VBM, the others lie
$\sim$0.12 eV and $\sim$0.37 eV higher in energy at $\Gamma$. They make their
minimum at $\Gamma$ and disperse up to the flat section along M--X$'$. Fermi
energy is set at the top of the upper lying state 0.46 eV above the VBM.
As in the case of Pt adsorption, bulk-like surface states fall in the gap from
the CB by 0.1 eV at $\Gamma$. Resulting band gap narrowing of 0.56 eV
corresponds to visible optical response. Such point defects at the
interstitial sites might also be important to get a better understanding of the
complex behavior of titania. For instance, in order to offer a possible
explanation for the appearance of Ti$3d$ defect state observed on (110) surface
reduced by the loss of bridging oxygen(s), Wendt \textit{et al.} considered Ti
atom at interstitial sites that yields a gap state as observed~\cite{wendt}.

\section{Conclusions}

Au atom binds to the stoichiometric rutile (110) surface much weaker than Pt
does. Theory suggests a three dimensional clustering upon gold deposition on
the defect-free surface. In this sense, a full Au adlayer is difficult to
realize whereas Pt coverage is more probable. Au and Pt diffusion into
the (1$\times$1) unit cell is thermodynamically unstable. However, their
substitution for Ti5c become even more stable than the clean surface at low
concentrations under oxygen rich conditions. Noble metal incorporated phases
on (3$\times$2) cell are found to be within the reach of thermal treatment.
Therefore, Pt and Au atoms can be adsorbed, or doped as substitutes for the
fivefold coordinated Ti atoms, and implanted into interstitial sites in the
lattice, at low concentrations. Formation of adsorptional impurities are
energetically more favorable than the other two phases.

Both noble metals are expected to promote the catalytic behavior of
TiO$_2$(110) surface by increasing the reaction probabilities through
availability of band--gap states. Stoichiometric TiO$_2$
gains metallic character upon single Au atom presence due to an unpaired
$5d$ electron. Moreover, narrowing of the gap towards visible region results from
impurity driven defect states for Pt which, hence, can be used in
photocatalysis. The wavelength tuning of photo response of titania might be
achieved by different types of Pt incorporation at low coverages.

\section*{Acknowledgments}
We acknowledge partial financial support from the Scientific and Technological
Research Council of Turkey (T\"{U}B\.{I}TAK) (Grant No: 110T394).

\end{document}